**Impact of stacking faults and domain boundaries on the electronic transport in cubic silicon carbide probed by conductive atomic force microscopy**

*F. Giannazzo[1*], G. Greco[1], S. Di Franco[1], P. Fiorenza[1], I. Deretzis[1], A. La Magna[1], C. Bongiorno[1], M. Zimbone[1], F. La Via[1], M. Zielinski[2], F. Roccaforte[1]*

Dr. F. Giannazzo, Dr. G. Greco, Dr. S. Di Franco, Dr. P. Fiorenza, Dr. I. Deretzis, Dr. A. La Magna, Dr. C. Bongiorno, Dr. M. Zimbone, Dr. F. La Via, Dr. F. Roccaforte

Consiglio Nazionale delle Ricerche – Istituto per la Microelettronica e Microsistemi (CNR-IMM)
Strada VIII, n.5
Zona Industriale, I-95121 Catania, Italy
e-mail: filippo.giannazzo@imm.cnr.it

Dr. M. Zielinski

NOVASiC,
Savoie Technolac,
BP267, F-73375
Le Bourget-du-Lac Cedex, France

**Abstract**

In spite of its great promises for energy efficient power conversion, the electronic quality of cubic silicon carbide (3C-SiC) on silicon is currently limited by the presence of a variety of extended defects in the heteroepitaxial material. However, the specific role of the different defects on the electronic transport is still under debate. In this work, a macro- and nano-scale characterization of Schottky contacts on 3C-SiC/Si was carried out, to elucidate the impact of the anti-phase-boundaries (APBs) and stacking-faults (SFs) on the forward and reverse current-voltage characteristics of these devices. Current mapping of 3C-SiC by conductive atomic force microscopy (CAFM) directly showed the role of APBs as the main defects responsible of the reverse bias leakage, while both APBs and SFs were shown to work as preferential current paths under forward polarization. Distinct differences between these two kinds of defects were also confirmed by electronic transport simulations of a front-to-back contacted SF and APB. These experimental and simulation results provide a picture of the role played by different types of extended defects on the electrical transport in vertical or quasi-vertical devices based on 3C-SiC/Si, and can serve as a guide for improving material quality by defects' engineering

**Introduction**

Silicon carbide (SiC) is an excellent semiconductor for the fabrication of power electronics devices [1]. Today, most of the fundamental research related to SiC and the devices development are focused on the hexagonal polytype (4H-SiC). In fact, 4H-SiC is already a mature material, currently available on large diameter (up to 150 mm) wafers of good electronic quality [2].

However, since some decades the cubic polytype (3C-SiC) has been attracting a notable scientific interest in the community [3,4,5,6,7]. In fact, due to its different band structure, 3C-SiC can give important technological advantages with respect to 4H-SiC. As an example, due to the lower band gap of the polytype (~2.3eV), a low density of interface states is expected at the $SiO_2$/3C-SiC interface close to the conduction band edge [8,9], which should result into a high inversion channel mobility in metal-oxide-semiconductor field effect transistors (MOSFETs) [10,11].

A unique feature of the 3C-SiC polytype is the possibility to be grown on large diameter silicon (Si) substrates. Hence, the 3C-SiC/Si systems are of great interest to integrate the superior properties of SiC into the well-established Si technology and achieve excellent performances at a lower cost than 4H-SiC. As an example, *Tanner et al.* [12] recently reported excellent rectifying properties of n-3C-SiC/p-Si heterojunctions, which can find application in high temperature electronics and optoelectronics devices.

In spite of this great potential, the use of 3C-SiC/Si for power devices is still hindered by the quality of the available material, which is characterized by the presence of a high density of extended defects, such as micro-twins, stacking faults (SFs) and crystalline domains boundaries, specifically the anti-phase boundaries (APBs) [7].

Achieving good Schottky contacts has been always a challenge for device fabrication on 3C-SiC. In fact, even high work function metals, like Au, typically exhibited low values of the Schottky barrier height on 3C-SiC compared to the theoretical predictions [13]. Hence, understanding the origin of the leakage current in relation to the material quality has been the object of scientific debate in the last years.

Using a nanoscale characterization by conductive atomic force microscopy (CAFM), *Eriksson et al.* [14] correlated the poor rectifying behavior of Au Schottky contacts on 3C-SiC layers grown onto on-axis 4H-SiC to the presence of extended defects (i.e., SFs, double position boundaries, triangular pits). Similarly, nanoscale electrical characterization of 3C-SiC layers grown on low off-axis 4H-SiC substrate suggested that conductive surface defects, probably associated to SFs, are responsible for the high leakage current and deviation from the ideal Schottky behavior [15]. More recently, correlating the leakage current in 3C-SiC/Si heterojunctions with the surface defects observed in the 3C-SiC layers suggested that APBs and SFs do not substantially contribute to the leakage phenomenon [16].

Clearly, this topic is still object of controversial interpretation and further experimental and theoretical investigations are required to get a clear understanding of the role played by extended defects in heteroepitaxial 3C-SiC on the electrical behavior of Schottky diodes fabricated on this system.

This paper presents a multiscale (i.e. macro- and nano-scale) characterization of Pt-Schottky contacts on 3C-SiC/Si, to elucidate the impact of APBs and SFs on the forward and reverse current-voltage characteristics of these devices. Statistical analysis on Schottky diodes with variable size showed a decrease of the forward bias turn-on voltage ($V_{on}$) and an increase of the reverse bias leakage current density with increasing the contact area. Furthermore, an areal density $D\approx2\times10^5 cm^{-2}$ of the defects responsible for the enhanced leakage current was evaluated from this statistical analysis. Nanoscale resolution current mapping of 3C-SiC by CAFM provided a direct demonstration of the key role of APBs as the main extended defects responsible for the enhanced leakage under reverse bias, while both APBs and SFs were shown to work as preferential current paths under forward polarization. Electronic transport simulations of a front-to-back contacted SF and APB showed a significantly increased transmittance as compared to defect-free 3C-SiC, and distinct differences between these two types of defects, thus confirming their key role as preferential current paths under particular bias polarizations.

## 2  Materials and Methods

Unintentionally doped 3C-SiC films with 10 μm thickness were grown by CVD on a n-type (1-10 Ω cm resistivity) on-axis Si(100) substrate, after the formation of a buffer layer by carbonization of the Si surface [17]. A schematic cross-section of the heteroepitaxial structure is illustrated in Fig.1(a). After the growth, a chemical mechanical polishing of the surface was carried out to reduce the roughness. Morphological analyses were carried out by atomic force microscopy (AFM) using a DI3100 equipment with Nanoscope V controller. Plan-view and cross-sectional transmission electron microscopy (TEM) images were acquired using a 200 kV JEOL 2010 F Microscope. Atomic resolution scanning transmission electron microscopy (STEM) images in cross-section were carried out with an aberration-corrected JEOL ARM200F microscope. Arrays of Pt Schottky diodes with variable size were fabricated on the 3C-SiC surface as follows. First, a $Ni_2Si$ Ohmic contact was obtained by deposition of a Ni frame on the sample front side, followed by thermal annealing at 950 °C [18]. Afterward, Pt contacts with circular shape and radius ranging from 5 to 25 μm were deposited inside this frame. Optical microscopy of the array of Pt contacts on 3C-SiC is reported in Fig.1(b).

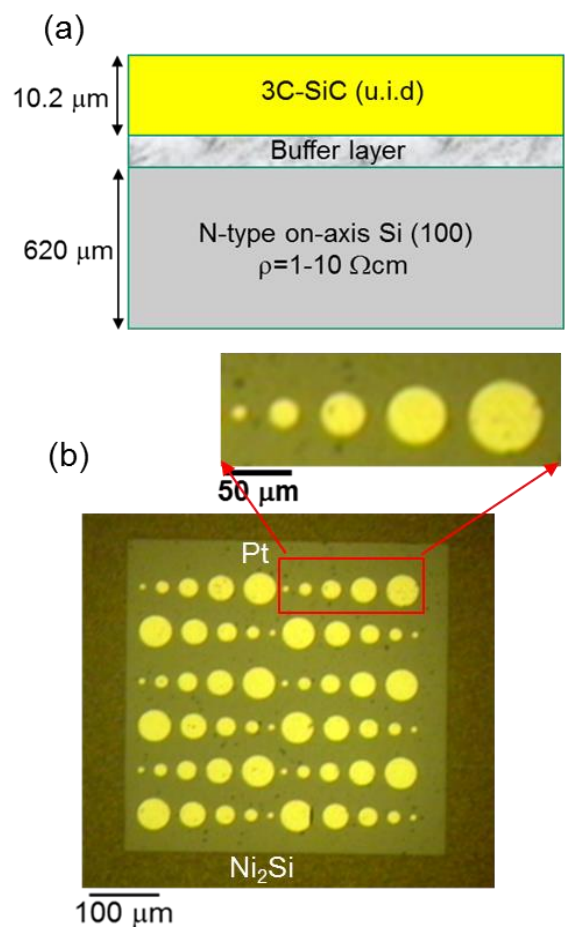

**Fig.1** (a) Schematic cross-section of the heteroepitaxial 3C-SiC/Si structure. (b) Optical microscopy of the array of Pt Schottky contacts on 3C-SiC and of the $Ni_2Si$ Ohmic contact.

Conductive Atomic Force Microscopy (CAFM) analyses were carried out using Pt coated Si tip connected to a current amplifier, by applying a potential difference between the tip and the $Ni_2Si$ Ohmic contact. Current maps were acquired on the bare 3C-SiC surface nearby the Pt contacts. Furthermore, the same CAFM tip was used to contact the Pt diodes with different areas and perform current-voltage measurements.

Density functional theory calculations were performed with the SIESTA code [19], using the local density approximation [20] for exchange and correlation. We calculated the band structure of defect-free, single SFs and single APBs, constructing appropriate supercells that minimize the interactions between neighboring defects in the periodic images. For the SF, we considered a hexagonal supercell along the [111] direction, whereas, for the APB, a rectangular supercell was been expanded in the [110] crystallographic direction. The wave functions were written on a double-ζ polarized basis set for both elements, while the ionic cores were described with norm-conserving Troulier-Martins pseudopotentials [21]. A mesh cutoff energy of 450 Ry was used for real-space integrals. All atoms were allowed to relax until forces were less than 0.04 eV/A˚. Quantum transport calculations were performed within the nonequilibrium Green's function formalism as described in Ref. [22]. In order to calculate the transport properties of front-to-back contacted SFs and APBs, appropriate rectangular supercells were constructed, where the defects were ideally contacted from the to the bottom electrode.

## 3   Results and discussion

Fig.2(a) schematically illustrates the setup for the electrical characterization of the variable size Pt diodes on 3C-SiC. Furthermore, two representative sets of current density vs bias (J-V) curves collected on arrays of diodes with 5 μm and 25 μm radius are reported in Fig.2(b) and (c), respectively. Although a rectifying behavior can be observed for both sets of measurements, the curves collected on the larger diodes (25 μm radius) showed a significantly higher leakage current density under reverse polarization, with a larger spread between different diodes, as compared to the measurements on the smaller (5 μm) contacts.

Since the reverse leakage current is a very relevant parameter for the diodes' performance, fixing a threshold current density and sorting the diodes with a leakage below this value is a reasonable criterion to estimate the "yield" of working devices. As illustrated by the horizontal dashed lines in Fig.2(b) and (c), a threshold current density of 10 μA/cm$^2$ has been fixed for our set of measurements. The experimental values of the yield versus the diode areas are reported by circles in Fig. 2(d). The red line is the fit of the data with the function:

$$Y(D) = \left(\frac{1-e^{-DA}}{DA}\right)^2$$

where A is the contact area and D is the areal density of defects responsible for devices failure. This probability function is commonly employed to describe the yield of electronic devices, and it has been also applied to the case of Schottky diodes onto 3C- on 4H-SiC substrates [14,23]. In the present case, the evaluated density $D=2\times10^5$ cm$^{-2}$ is associated with the extended defects responsible for a leakage current exceeding the fixed threshold of 10 µA/cm$^2$. From this areal density value, a typical distance $L=1/D^{1/2}\approx20$ µm between these defects can be estimated.

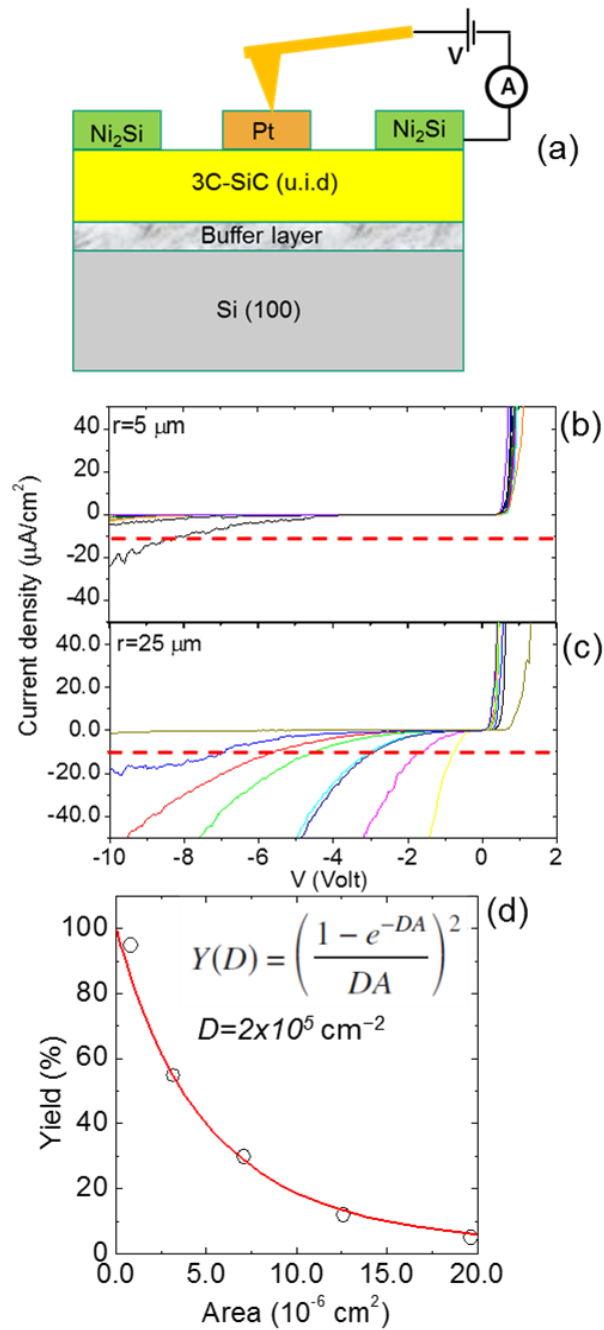

**Fig.2** (a) Schematic illustration of the setup for the electrical characterization of the variable size Pt diodes on 3C-SiC. Sets of J-V curves collected on arrays of diodes with (b) 5 μm and (c) 25 μm radius. (d) Percentage of diodes (yield) with a reverse leakage lower than a threshold current density of 10 μA/cm$^2$, as a function diodes area. The red line is the fit with the function reported in the insert and D is the areal density of defects responsible for the leakage.

The Schottky contact area was found to have an impact not only on the reverse bias leakage current but also on the current onset under forward bias polarization. Fig.3 shows the forward bias J-V characteristics measured on the diodes with 5 μm (a) and 25 μm (b) radius in a bias range from 0 to 1.2 V. For each curve, the turn-on voltage ($V_{on}$) was evaluated as the bias corresponding to a current density of 0.1 μA/cm$^2$. The resulting histograms of $V_{on}$ for the two sets of diodes with 5 and 25 μm radius are shown in the right panels of Fig. 3(a) and (b), respectively. The average value of $V_{on}$ was found to increase from ~0.12 V (for r=25 μm) to ~0.4 V (for r=5 μm). This $V_{on}$ value approximately corresponds to the Schottky barrier $\Phi_B$ of the Pt/3C-SiC contact. In this regard, it is worth noting that the obtained experimental $V_{on}$ values are much lower than the theoretical value for the ideal Pt/3C-SiC Schottky barrier, obtained according to the Schottky-Mott model, i.e., $\Phi_B=W_{Pt}-\chi_{3C-SiC}\approx 1.6$ eV, where $W_{Pt}=5.6$ eV is the Pt workfunction and $\chi_{3C-SiC}=4$ eV is the 3C-SiC electron affinity. Such very low values of $V_{on}$ can be ascribed to the presence of conductive paths in the 3C-SiC material, giving rise to an enhanced forward current injection with respect to that expected for an ideal Pt Schottky contact on this wide bandgap semiconductor.

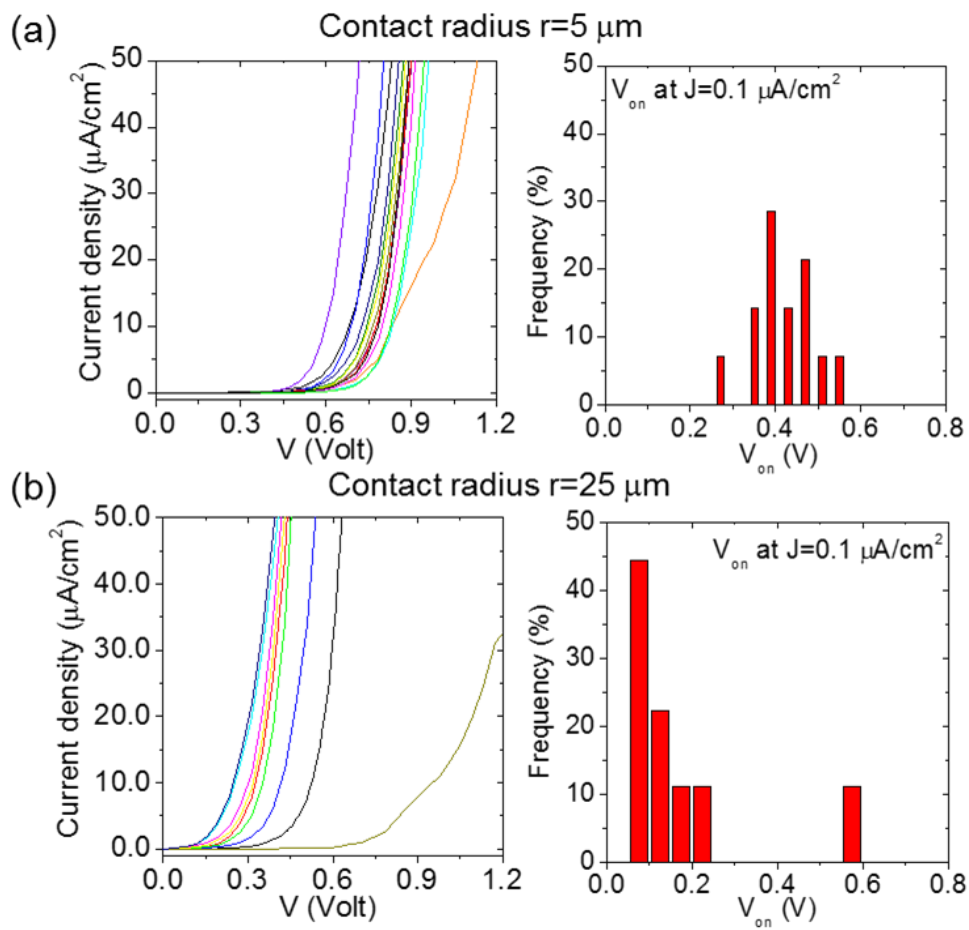

**Fig.3** Forward bias J-V characteristics measured on the diodes with (a) 5 μm and (d) 25 μm radius. Histograms of the turn-on voltage ($V_{on}$) for the two sets of diodes are shown in the right panels of (a) (b), respectively.

In order to elucidate the nature of the defects responsible of the enhanced reverse leakage current and of the low turn-on voltage observed in the Schottky diodes, a nanoscale resolution structural and electrical characterization of the 3C-SiC layer on Si has been carried out.

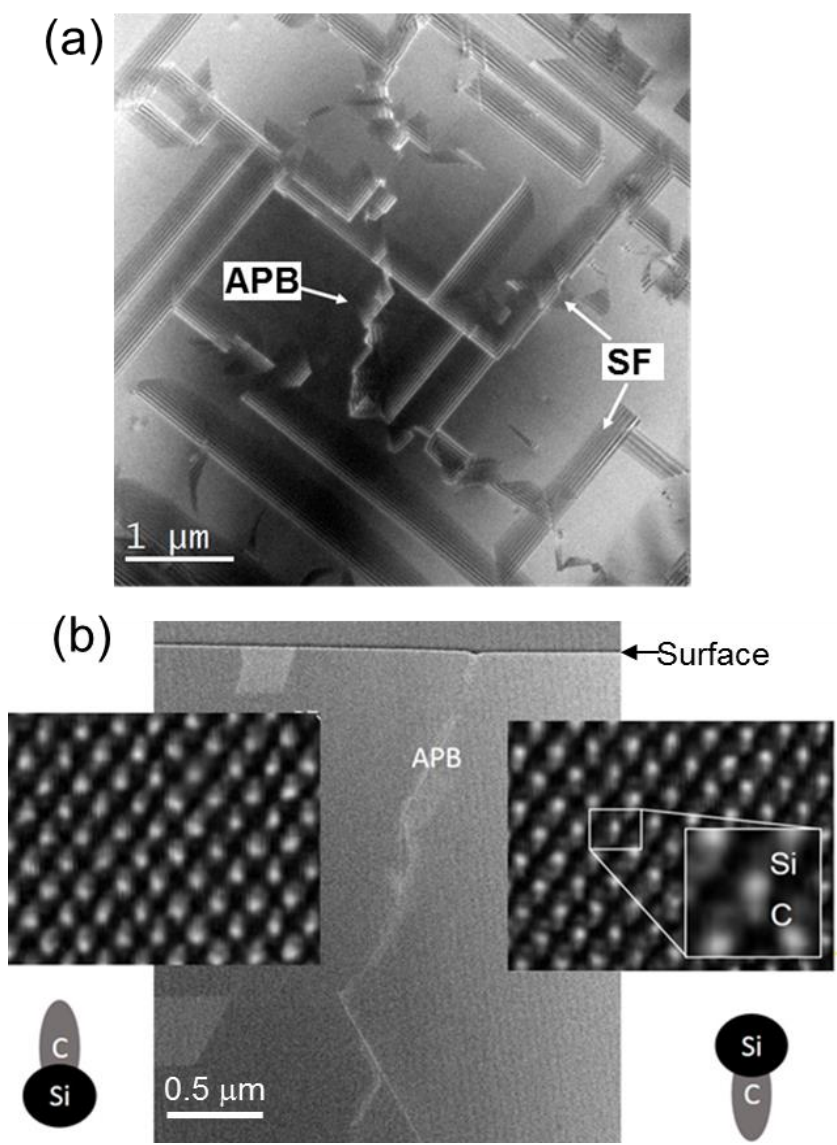

**Fig.4** (a) Plan-view TEM image collected on a 3C-SiC lamella thinned from the backside to remove the substrate. (b) Cross-sectional TEM image of the 3C-SiC layer with a domain boundary. Left and right inserts show two atomic resolution STEM analyses of the 3C-SiC domains at the two sides of the domain boundary. The inversion of crystal symmetry moving from one crystalline domain to the other demonstrates the APB nature of the domain boundary.

Fig.4(a) shows a typical plan-view transmission electron microscopy (TEM) collected on a 3C-SiC lamella thinned from the backside to remove the substrate. The presence of a large number of SFs (i.e. planar crystallographic defects lying on {111} planes) and of a domain boundary can be observed in the 5μm ×5μm imaged area. Detailed structural analysis by atomic resolution cross-sectional analysis revealed that these domain boundaries are antiphase boundaries (APBs) i.e. boundaries between domains that are rotated upside down. Fig.4(b) reports a representative low magnification cross-sectional STEM image of the 3C-SiC layer, showing a domain boundary which extends in all the imaged thickness up to the surface, where it introduces a few nm "V-shape" depression. Two atomic-resolution STEM analyses of the 3C-SiC domains at the two sides of the domain boundary are also reported in the left and right inserts of the image. An inversion of crystal symmetry is clearly observed, with the upside-down flipping of the Si-C bond moving from one crystalline domain to the other. It demonstrates the APB character of the domain boundary.

Nanoscale resolution current mapping of the bare 3C-SiC surface was carried out by CAFM, as schematically depicted in Fig.5(a). A representative morphological image collected on a 20 μm × 20 μm scan area is reported in Fig.5(b), from which the surface root mean square (RMS) roughness of 3.2 nm was evaluated. An APB can be easily identified as a nanometer deep "V-shape" depression in the topographic map, as well as in the height line-scan reported in Fig.5(b), right panel. Fig.5(c) and (d) show the current maps measured simultaneously to the topography by reverse bias ($V_{tip}$=-0.5 V) and forward bias ($V_{tip}$=0.5 V) polarization of the Pt tip, respectively. The current values measured under reverse polarization are much lower than those measured under forward polarization, thus confirming the Schottky diode behavior of the Pt/3C-SiC contact also at nanoscale level. Using the same current range (from 0 to 50 pA) for the two current maps, APBs are the most evident conductive features under reverse bias, whereas both APBs and SFs (indicated by blue arrows in Fig.5(d)) contribute to the conduction under forward polarization. Two representative scan lines across the APB for the two opposite tip biases are also reported in the right panels of Fig.5(c) and (d), showing a more than 10 times higher current peak on the APB under forward bias with respect to the reverse one. These results suggest that APBs are the main responsible of the enhanced reverse leakage current measured in macroscopic Pt/3C-SiC Schottky diodes. Noteworthy, the separation between these extended defects deduced from this microscopic analysis is in the order of tens of micrometers, in very close agreement with the value of $L$ (≈20 μm) deduced from the statistical characterization of the diodes.

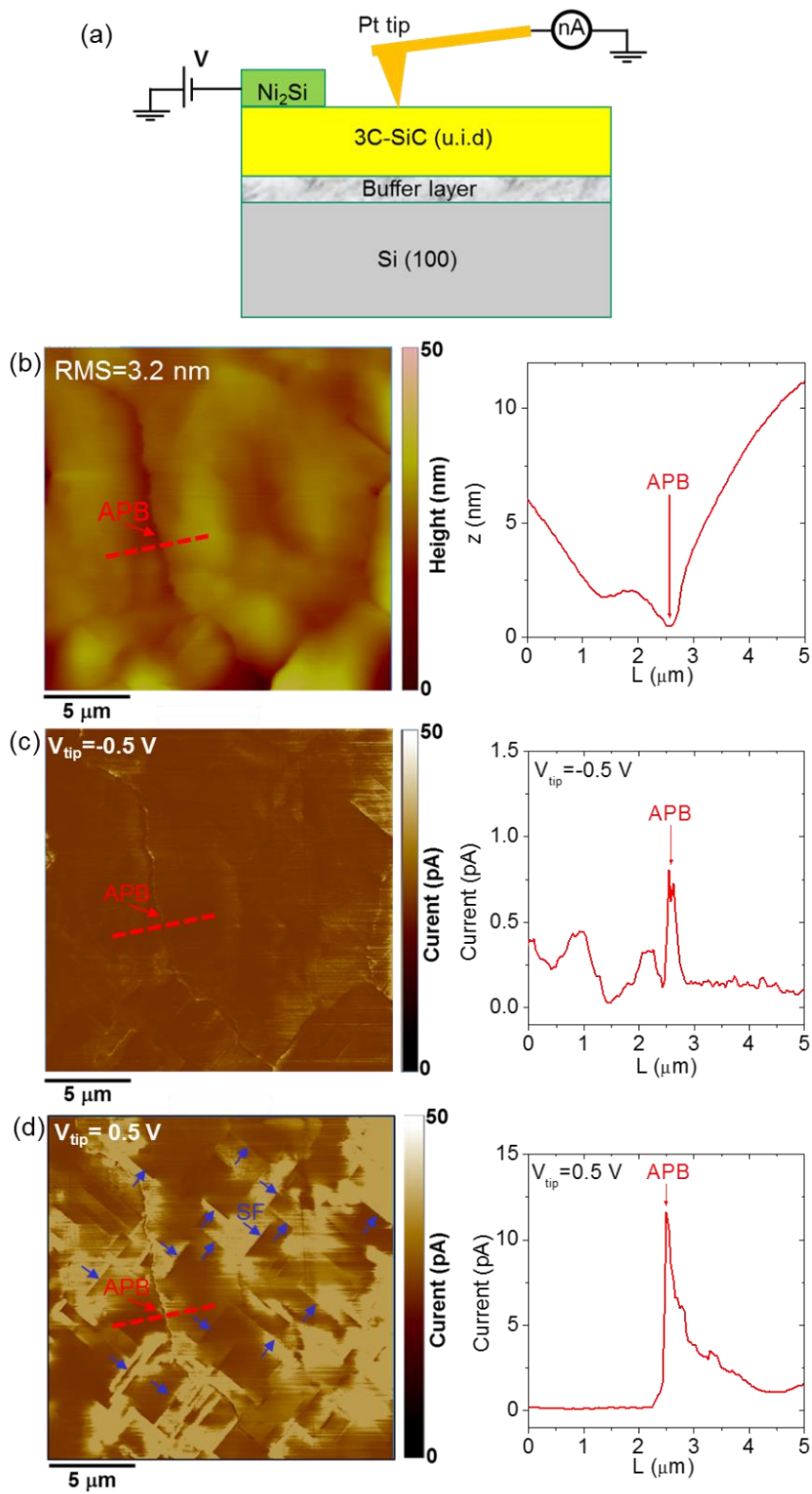

**Fig.5** (a) Schematic illustration of the CAFM setup. (b) Morphology and current maps collected (c) under reverse bias polarization of the tip ($V_{tip}$=-0.5 V) and (d) under forward bias polarization ($V_{tip}$=0.5V). An APB is indicated by a red arrow and SFs by blue arrows, respectively. Representative line-scans across a grain boundary extracted from the topography (b, right panel), current maps under reverse bias polarization (c, right panel) and forward bias polarization (d, right panel) of the tip are shown.

Fig.6(a) shows a zoom-in of the forward bias current map inside the 3C-SiC grain, showing a more detailed view of SFs. In most of the cases, these extended defects can be identified as sharp lines, which separate a highly conductive region (indicated by "1") and a lowly conductive one (indicated by "2") in the current map. This peculiar current contrast is a consequence of the inclination of the SF plane {111} with respect to the [100] growth axis. Fig.6(b) schematically illustrates the forward biased conductive tip scanned on the 3C-SiC surface across a SF. When the tip is placed on the side labeled with "1", the electrons overcoming the Pt/3C-SiC Schottky barrier are channeled in the SF conductive path. On the other hand, when the tip is placed on the opposite side of the SF (labeled with "2"), the injected electrons will travel through the defect-free 3C-SiC epitaxy, resulting in a smaller current.

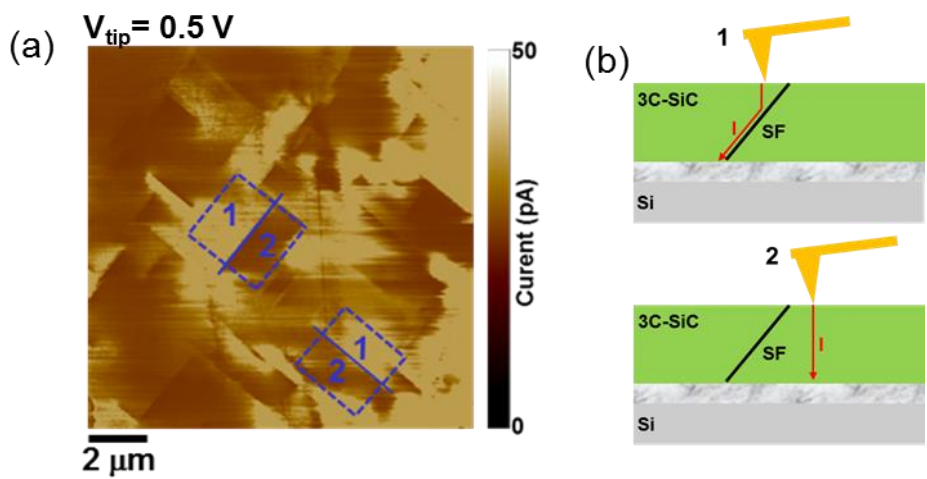

**Fig.6** (a) Current map collected under forward polarization of the Pt tip ($V_{tip}$=0.5 V), showing a more detailed view of stacking faults (SF) areas. (b) Schematic illustration of the origin of different current contrast at the two sides of SFs in the CAFM map.

Finally, quantum transport calculations have been carried out, in order to elucidate the origin of the enhanced conductivity of 3C-SiC APBs under reverse and forward polarization, and the preferential conductivity of SFs under forward polarization. The calculated transmission coefficient versus the energy and wave-vector for defect-free 3C-SiC and 3C-SiC with a single SF and an APB is reported in Fig.7, considering different contact geometries. Fig.7(a) illustrates the case of a defect-free 3C-SiC layer which is laterally contacted by two electrodes, resulting in a current flow in the direction [110] perpendicular to the growth axis [100]. Fig.7(b) illustrates the results obtained in the configuration corresponding to the experimental setup employed in our CAFM investigations, i.e. an SF contacted from a top electrode (the tip) and a back electrode (corresponding to the Si substrate). A significant enhancement of the transmittance both in the valence and the conduction band of 3C-SiC can be observed in presence of the SF. Lets consider, however, that such increased conductivity does not impact on the transport within the 3C-SiC bandgap, since SFs in 3C-SiC do not introduce energy levels within the band-gap. In this sense, SFs can be considered as highly conducting 2D defects, but only within the energy range where also the bulk material is conductive. This is not the case of the APB defects (and in general the domain boundaries), which on the contrary introduce states within the 3C-SiC bandgap at the vicinity of the valence band. This aspect reduces the transport gap when an APB is front-to-back conducted (Fig. 7(c)), allowing the conduction of current even at energies that lie within the 3C-SiC bandgap. This last feature supports our observation of an increased conductance of APBs under forward and reverse polarization regimes of the tip.

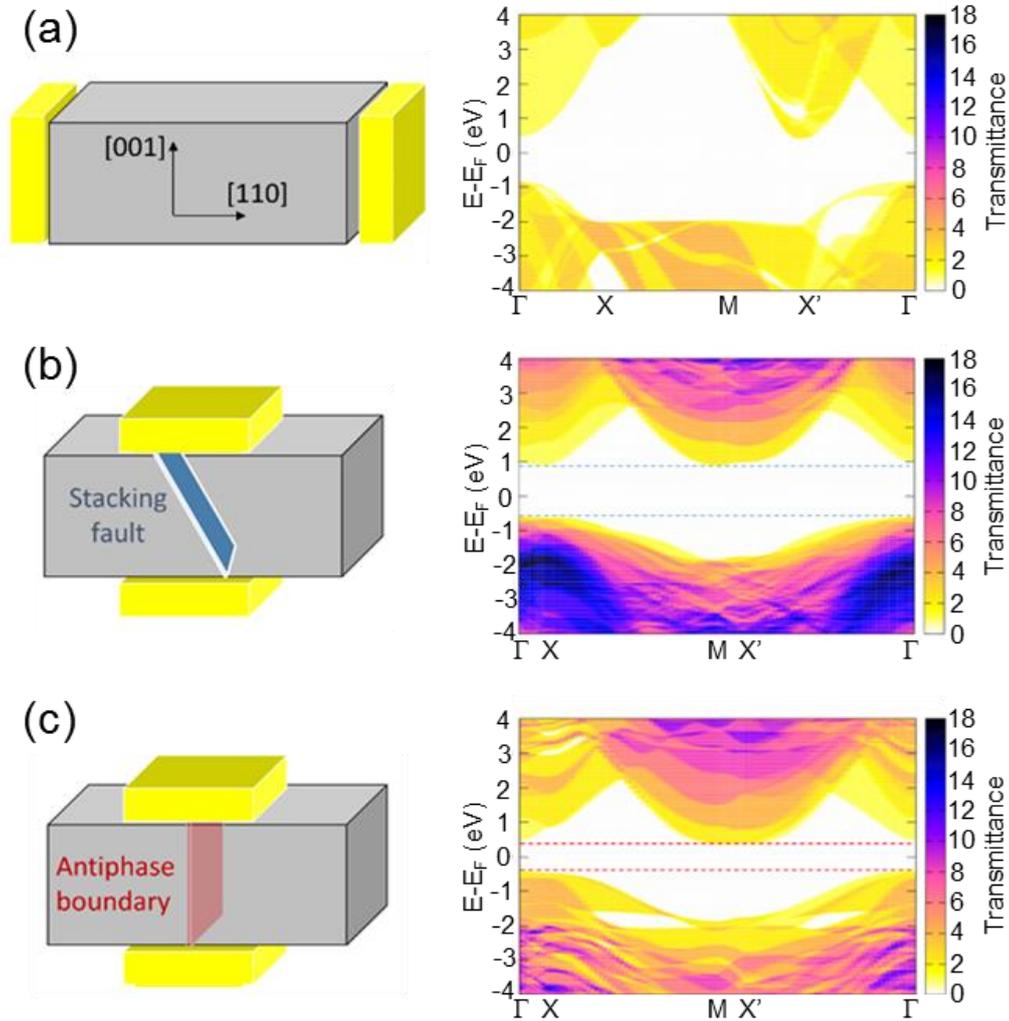

**Fig.7** Calculated transmission coefficient (seen as the color scale values) as a function of the electron energy and wave-vector for defect-free 3C-SiC and 3C-SiC with a single stacking fault (SF) and an anti-phase-boundary (APB): (a) laterally contacted defect-free 3C. (b) Front to back contacted 3C-SiC with an SF quasi-parallel to the current flow. (c) Front to back contacted 3C-SiC with an APB parallel to the current flow.

## 4      Conclusion

In conclusion, the role of characteristic extended defects (APBs and SFs) on the electronic transport in heteroepitaxial 3C-SiC onto Si has been clarified by the combination of current-voltage analyses on Pt/3C-SiC Schottky diodes, on nanoscale current mapping of 3C-SiC (with CAFM), and confirmed by electronic transport calculations. APBs were demonstrated to be the main responsible for the enhanced leakage current of Schottky diodes under reverse bias polarization. On the other hand, both APBs and SFs were shown to work as preferential current paths responsible for the reduced turn-on voltage under forward polarization. Electronic transport simulations of a front-to-back contacted SFs and APB demonstrated an increased transmittance as compared to the case of defects' free 3C-SiC, but with distinct differences between these two kinds of defects. Indeed, the presence of SFs results in an enhanced transmittance in the valence and conduction band of 3C-SiC without changes in the forbidden band-gap, whereas a significant shrinkage of the transport gap was found in the case of the APBs. These experimental and simulation results provided an insight of the electrical transport phenomena in vertical or quasi-vertical devices based on 3C-SiC/Si, and can serve as a guide for improving the heteroepitaxial material by defects' engineering.

**Acknowledgements**

M. Spera and C. Calabretta, PhD students at CNR-IMM in Catania, are acknowledged for their participation in devices electrical characterization and TEM samples preparation, respectively. This work has been supported by the H2020 project Challenge, Grant agreement n. 720827.

Submitted: October 22, 2019